\documentclass{article}
\usepackage{srcltx}
\usepackage{amsmath}
\usepackage{amssymb}
\usepackage{amsfonts}
\usepackage{latexsym}
\usepackage{textcomp}
\usepackage{appendix}
\usepackage{multirow}
\usepackage{booktabs}
\usepackage{url}
\usepackage{array}
\usepackage{marvosym}
\usepackage{graphics}
\usepackage{graphicx}
\usepackage{subfigure}
\usepackage{epsfig}

\usepackage{color}
\usepackage{multicol}        
\usepackage[bottom]{footmisc}

\title{Stock returns forecast: an examination by means of Artificial Neural Networks}

\author{Mart\'in Iglesias Caride \\
\footnotesize Master Program in Data Mining, Universidad de Buenos Aires (Argentina) \\
Aurelio F. Bariviera \\
\footnotesize Department of Business, Universitat Rovira i Virgili, Av. Universitat 1, 43204 Reus (Spain) \\ \footnotesize \texttt{aurelio.fernandez@urv.cat} \\
 Laura Lanzarini \\
 \footnotesize Instituto de Investigaci\'on en Inform\'atica LIDI \\
 \footnotesize Facultad  de Inform\'atica, Universidad Nacional de La Plata (Argentina)
}
 
\begin{document}

\maketitle

\begin{abstract}
The validity of the Efficient Market Hypothesis has been under severe scrutiny since several decades. However, the evidence against it is not conclusive. Artificial Neural Networks provide a model-free means to analize the prediction power of past returns on current returns. This chapter analizes the predictability in the intraday Brazilian stock market using a backpropagation Artificial Neural Network. We selected 20 stocks from Bovespa index, according to different market capitalization, as a proxy for stock size. We find that predictability is related to capitalization. In particular, larger stocks are less predictable than smaller ones. 
\end{abstract}%

\section{Introduction}
\label{sec:intro}
The Efficient Market Hypothesis (HME) is a theoretical framework, which constitutes the baseline of financial economics. It was developed following the guidelines of neoclassical economics \cite{ROss05}. According to the standard definition by Fama \cite{FAma70}, a market is said to be informational efficient if prices conveys all relevant information. The definition immediately leads to the determination of the relevant information set. For this reason, Fama \cite{FAma70} classifies informative efficiency into three categories: (i) Weak efficiency, when the current price contains information from the past series of prices; (ii) Semi-strong efficiency, when the past price contains all the public information associated with that asset; and (iii) strong efficiency, when the price reflects all public and private information relating to that asset. HME has been the unquestioned paradigm in economics until the 1980s. With the advent of personal computers and increased availability of data, HME testing became easier. Thus, in those years began to appear articles questioning the weak version of informational efficiency. The first articles inquired about seasonal effects on returns. Later, researchers began to look at return forecasts. In a review of his 1970 article, Fama \cite{FAma91} renames the first category of weak efficiency as \textit{tests for return predictability}, where explanatory variables, in addition to past performance may be other financial variables such as interest rates, price-earning ratio , etc. If EMH is an adequate description of market behavior, the forecast of returns (whether based on past returns or adding other variables) is ruled out. In particular, all chartist theories that advocate the existence of more or less fixed patterns in the financial markets are discarded. 

The aim of our paper is to test the EMH in its weak version for the Brazilian stock market, using sign prediction (upward or downward movement) through an Artificial Neural Network (ANN).

The rest of the paper is organized as follows: Section \ref{sec:ANN} makes a brief review of the use of ANN in economics; Section \ref{sec:methodology} describes the methodology; Section \ref{sec:data} describes data used in the empirical  application carried out in Section \ref{sec:results}. Finally, Section \ref{sec:conclusions} draws the main conclusions of our research.

\section{Stock market forecasts: Artificial Neural Networks \label{sec:ANN}}
The use of ANNs to economic and financial problems is very prolific. ANNs, depending on their architecture and type, can be used to classify, to optimize and to forecast. This versatility, including their robustness and ability to deal with nonlinear dynamics, is undoubtedly part of their success. Usually, many business problems involve several variables, with an unknown functional (nonlinear) relationship. This
 situation enhances the use of data driven models that can be model using ANNs. 
 
G{\"{o}}{\c{c}}ken \textit{et al.} \cite{Gocken2016} propose an hybrid model consisting in a harmony search or 
a genetic algorithm in combination with an ANN, in order to enhance the capability of return forecasts in the Turkish stock market.
An hybrid approach was also selected by Qiu \textit{et al.} \cite{Qiu2016}. They combine the ANN with a genetic algorithm or a simulated annealing, in order to filter the variables to be set in the input layer. 

In addition to being used in stock market forecasting, ANNs are used also in credit scoring. For example, \cite{Lanzarini2015} use and hybrid approach, that combines  a neural network with an optimization technique, in order to classify customers of two public databases. Although results are rather similar in terms of accuracy, this novel approach allows to simplify the number of rules to obtain a given level of accuracy. The same technique was extended in \cite{Lanzarinietal2017} to a real and large database of customers of a financial institution in Ecuador. The outcome was a more intuitive customer classification, which helps managers to know with greater detail which variables really matters in a client credit application. 
 
A recent survey by \cite{Tkac2016}, finds 412 articles in leading academic journals that apply ANN to business and economics. The survey covers the last two decades. The applications ranges from credit scoring, to stock forecast, to marketing. This is the reflect of an active research area, which applies data science to add value to business.

\section{Methodology \label{sec:methodology}}
Lanzarini \textit{et al.} \cite{Lanzarini2011} studies the prediction ability of a backpropagation ANN, using daily values of an stock index. In this paper, instead of working with daily index values, we will work with high frequency intraday stock prices. We would like to test the ability of an ANN to forecast movements in times of minutes. The rationale for this study is the following: many investing firms perform \textit{algorithmic trading},  i.e. firms operate very frequently in markets and their buy and sell decisions are made through machine learning algorithms. Consequently, it is an important topic for the financial industry. 

The prediction was solved by means of an ANN \cite{Isasi04}. This kind of construct tend to simulate the way human brain learns, by experience. Basically we can think of a neural network as a directed graph whose nodes are called neurons and the lines that connect them have an associated weight. This value in some sense represents the ``knowledge'' acquired. Neurons are organized into layers according to their function: the input layer provides information to the network, the output layer provides the answer and the hidden layers are responsible for carrying out the mapping between input and output \cite{Freeman91}.

In this paper we use a multilayered neural network. It is a feedforwad network, totally connected, organized in three layers: 10 input neurons, six neurons in a single hidden layer and one output neuron. This architecture coindices with the one used in \cite{Gencay99},\cite{Lanzarini2011} and \cite{FErro00}. 

Differently the learning algorithm is a resilient propagation instead of the more common backpropagation. The training is supervised. The resilient propagation otherwise of gradient descent performed by the backpropation updates each weight independently, and is not subject to the size of the derivative influence but only dependant on the temporal behaviour of its sign (\cite{Riedmiller93adirect}). The network was designed to forecast the market return. 

Let $P=(p_{1}, p_{2}, ..., p_{L})$ be the sequence of stock quotes of figure \ref{fig:quotes}, then the instantaneous return at the moment $t+1$ is computed as indicated in (\ref{eq:equation1}): 
\begin{equation}
r_{t+1}=\ln \left( \frac{p_{t+1}}{p_{t}} \right)
\label{eq:equation1}
\end{equation}
where $p_{t+1}$ and $p_{t}$ are stock quotes at period $t+1$ and $t$, and $r_{t+1}$ is the continuous compound return obtained for buying in period $t$ and selling in period $t+1$. 

In order to help to improve network performance the trend of last nine returns which are used as input for the networks was calculated using ordinary least squared and used as the 10th input for the model. The fact that the training is supervised, implies knowing the expected value for each of the examples that will be used in the training.

Thus, it is required a set of ordered pairs $\{(X_{1}, Y_{1}), (X_{2}, Y_{2}), \ldots , (X_{j}, Y_{j}), \ldots, (X_{M}, Y_{M})\}$ being $X_{j}=(x_{j,1}, x_{j,2}, \ldots, x_{j,10})$ the input vector, from which $x_{j,10}$ is the trend of last nine returns and $Y_{j}$ the answer value that it is expected that the network learns for that vector. In this case:
\begin{equation}
x_{j,k}=r_{j+k}=\ln\left(\frac{p_{j+k}}{p_{j+k-1}}\right) \hspace{0.5cm} k=1, 2,\dots9
\label{eq:entrada1}
\end{equation}
\begin{equation}
Y_{j}=r_{j+1+N}=\ln\left(\frac{p_{j+1+N}}{p_{j+N}}\right)
\label{eq:entrada}
\end{equation}
The maximum number of pairs $(X_{j}, Y_{j})$ that can be formed with $L$ stock quotes is $M = L - 9$. The first half of them were used to train the network and the second half to verify its performance.
Once the network is trained, its answer for vector $X_{j}$ is computed as indicated in equation (\ref{eq:salida}).
\begin{equation}
Y^{\prime}_{j}=G \left( a_{0}+\sum_{i=1}^{6} a_{i} F \left( b_{0,i}+\sum_{k=1}^{10}x_{j,k} b_{k,i} \right ) \right)
\label{eq:salida}
\end{equation}
where $x_{j,k}$ is the value corresponding to the $k-th$ input defined in (\ref{eq:entrada1}), $b_{k,i}$ is the weight of the arch that links the $k-th$ neuron of the input with the $i-th$ hidden neuron and $a_{i}$ is the weight of the arch that links the $i-th$ hidden neuron with the single output neuron of the network. We should note that each hidden neuron  has an additional arch, whose value is indicated in $b_{0,i}$. This value is known as \textit{bias} or trend term. A similar thing happens to the output neuron with the weight $a_{0}$.

Finally, the obtained expected return $Y^{\prime}_{j}$  will be used to do the corresponding forecast. Instead of forecasting any positive (negative )value as a indicative that the return will positive (negative) and the stock should be bought (sold), a range was used. In case the predicted rate is greater than $X$, the stock should be bought. If the predicted rate is lower than $-X$, the stock should be sold. In case the predicted rate lays within this range, the prior position should be maintained. In case last decision was to buy, the rule dictates to keep the stock. On the contrary if last prediction was that the rate was going to decrease the short position should be maintained. Instead of using a fix range to determine where to buy or sell, the range was optimized during training.

\subsection{Profitability measures}
In order to measure the profitablity of the ANN and our benchmark model (\textit{naive} buy-and-hold), we selected some common metrics.
The profitability measures used on pairs $(X_{j}, Y_{j})$ of the training set are the following:
\begin{itemize}
\item{\textbf{ANN Buy$\&$Hold Train}: is the return obtained by a simple \textit{buy $\&$ hold} strategy and is computed by adding all the expected returns of the training set as indicated in (\ref{eq:bh_train})
\begin{equation}
bhTrain = \sum_{j=1}^{M/2}Y_{j}
\label{eq:bh_train}
\end{equation}
}
\item{\textbf{ANN Buy$\&$Hold Test}: is equivalent to the previous measure but applied to the testing set.
\begin{equation}
bhTest = \sum_{j=1+M/2}^{M}Y_{j}
\label{eq:bh_test}
\end{equation}
} 
\item{ \textbf{Sign prediction ratio}:  For rates sign predicted correctly we assign the value $1$ whereas if the sign was not predicted correctly the  $-1$ value corresponds.}
\begin{equation}
SPR = \frac{\sum_{j=1+M/2}^{M}matches(Y_{j},Y^{\prime}_{j})}{M/2}
\label{eq:aciertos}
\end{equation}

\begin{equation}
matches(Y_{j}, Y^{\prime}_{j}) = \left \{ \begin{array} {ll}
             1 &  if \  sign(Y_{j})=sign(Y^{\prime}_{j}) \\
             0 & if \ not
             \end{array}
            \right. 
\label{eq:iguales}
\end{equation}
where $sign$ is the sign function that maps $+1$ when the argument is positive and $-1$ when the argument is negative.
\item{The \textbf{maximum return}  is obtained by adding all the expected values in absolute value}
\begin{equation}
MaxReturn = \sum_{j=1+M/2}^{M}abs(Y_{j})
\label{eq:MaxReturn}
\end{equation}
and represents the maximum achievable return, assuming perfect forecast.
\item {The \textbf{total return} is computed in the following way
\begin{equation}
TotalReturn = \sum_{j=1+M/2}^{M}sign(Y^{\prime}_{j}) * Y_{j}
\label{eq:TotalReturn}
\end{equation}

where $sign$ is the sign function. Notice that the better the network prediction the larger the total return.}
\item{\textbf{Ideal Profit Ratio}  is the ratio between the total return (\ref{eq:TotalReturn}) and the maximum return (\ref{eq:MaxReturn}).
\begin{equation}
IPR = \frac{TotalReturn}{MaxReturn}
\label{eq:IPR}
\end{equation}
}
\item{\textbf{Sharpe Ratio} is the ratio between the total return and its standard deviation. The different values of the total return arises from the several independent runs in the training of the network. This will give different networks that will generate different total returns.
\begin{equation}
SR = \frac{\mu_{TotalReturn}}{\sigma_{TotalReturn}}
\label{eq:SR}
\end{equation}
}
\end{itemize}

\section{Data \label{sec:data}}

\begin{figure}
\includegraphics[scale=.55]{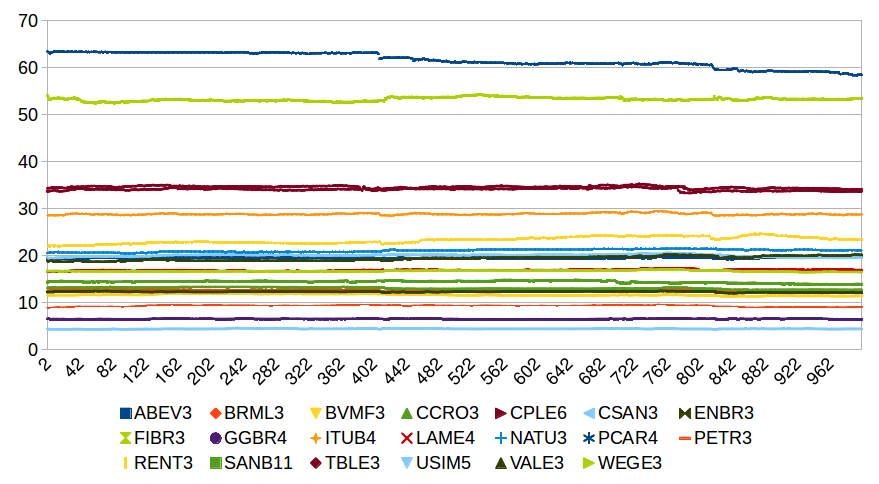}
\caption{Real time quotes for the twenty selected Bovespa's stocks.}
\label{fig:quotes}  
\end{figure}

We used real time quotes for a subset of 20 stocks which are included within the Bovespa Index. These stocks were selected based on their market capitalization and are detailed in Table \ref{tab:data}. The stocks which compose Bovespa were divided in quartiles and for each quartile the five stock with highest capitalization were selected in order to have a representative sample for each group.
For each stock one thousand data points were used from 09-16-2015 at 2:00 pm and as late as 09-18-2015 08:16 pm in order to restrict the running time of the algorithms used to build the models used. Figure \ref{fig:quotes} shows the quotes' evolution during the sample period.

\begin{table}
\caption{Details of stocks in the sample}
\begin{center}
\resizebox{\textwidth}{!}{
\begin{tabular}{rrrrr}
\toprule
Company Name & Sector & Market Capitalization  & Percentile & Ticker\tabularnewline
                           &            & (thousand USD) &  & \tabularnewline
\midrule 
Ambev SA & Consumer Staples & 93,084,295 & 75th \textendash{} 100th & ABEV3 BS Equity\tabularnewline
Itau Unibanco Holding SA & Financials & 61,371,355 & 75th \textendash{} 100th & ITUB4 BS Equity\tabularnewline
Petroleo Brasileiro SA & Energy & 56,071,359 & 75th \textendash{} 100th & PETR3 BS Equity\tabularnewline
Banco Santander Brasil SA & Financials & 25,954,473 & 75th \textendash{} 100th & SANB11 BS Equity\tabularnewline
Vale SA & Materials & 25,871,589 & 75th \textendash{} 100th & VALE3 BS Equity\tabularnewline
BMFBovespa SA & Financials & 10,031,988 & 50th \textendash{} 75th & BVMF3 BS Equity\tabularnewline
CCR SA & Industrials & 9,295,719 & 50th \textendash{} 75th & CCRO3 BS Equity\tabularnewline
WEG SA & Industrials & 8,235,428 & 50th \textendash{} 75th & WEGE3 BS Equity\tabularnewline
Engie Brasil Energia SA & Utilities & 7,856,355 & 50th \textendash{} 75th & TBLE3 BS Equity\tabularnewline
Lojas Americanas SA & Consumer Discretionary & 7,524,872 & 50th \textendash{} 75th & LAME4 BS Equity\tabularnewline
Cosan SA Industria e Comercio & Energy & 4,727,773 & 25th \textendash{} 50th & CSAN3 BS Equity\tabularnewline
Gerdau SA & Materials & 4,353,258 & 25th \textendash{} 50th & GGBR4 BS Equity\tabularnewline
Natura Cosmeticos SA & Consumer Staples & 4,146,966 & 25th \textendash{} 50th & NATU3 BS Equity\tabularnewline
Cia Brasileira de Distribuicao & Consumer Staples & 4,027,431 & 25th \textendash{} 50th & PCAR4 BS Equity\tabularnewline
Fibria Celulose SA & Materials & 3,762,646 & 25th \textendash{} 50th & FIBR3 BS Equity\tabularnewline
EDP - Energias do Brasil SA & Utilities & 2,678,135 & 0 \textendash{} 25th & ENBR3 BS Equity\tabularnewline
Localiza Rent a Car SA & Industrials & 2,634,754 & 0 \textendash{} 25th & RENT3 BS Equity\tabularnewline
BR Malls Participacoes SA & Real State & 2,312,292 & 0 \textendash{} 25th & BRML3 BS Equity\tabularnewline
Cia Paranaense de Energia & Utilities & 2,264,843 & 0 \textendash{} 25th & CPLE6 BS Equity\tabularnewline
Usinas Sider. de Minas Gerais SA & Materials & 2,164,639 & 0 \textendash{} 25th & USIM5 BS Equity\tabularnewline
\bottomrule 
\end{tabular}}
\label{tab:data}
\end{center}
\end{table}

\section{Empirical results \label{sec:results}}
We performed 30 independent runs for each stock in the sample described in Section \ref{sec:data}. The maximum number of iterations was capped at 3000. The initials weight were randomly distributed between $-1$ and $1$. Initial update value was set to 0.01, while the updates values were limited within $e^{-6}$ and $50$. Increase factor and decrease factor for weight updates was set to $1.2$ and $0.5$. The functions $F$ and $G$ used by the neural network are those defined in (\ref{eq:logsig}) and (\ref{eq:tansig}). Both are sigmoidal functions. The first one bounded between $0$ and $1$, and the second one is bounded between $-1$ and $1$. In this way, $F$ allows hidden neurons to produce small values (between 0 and 1), and $G$ permits the net to split between negative (expecting negative returns) and positive (expecting positive returns). 

\begin{equation}
G(n) = \frac{1}{1+e^{-n}}
\label{eq:logsig}
\end{equation}

\begin{equation}
F(n) = \frac{2}{1+e^{-2n}}-1
\label{eq:tansig}
\end{equation}

\begin{figure}[b]
\includegraphics[scale=.65]{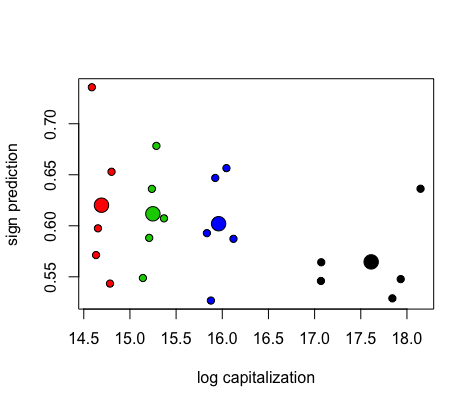}
\caption{Relationship between log market capitalization and sign prediction. Colors represent stocks of the same quartile. Bigger color dots reflect the sample mean of each quartile.}
\label{fig:2}  
\end{figure}

We show the results obtained for each stock in Tables \ref{tab:q1}, \ref{tab:q2}, \ref{tab:q3}, and \ref{tab:q4}, one for each stock quartile - by averaging the 30 runs and showing below the standard deviation when applicable. The sign prediction is always greater than 50\%. We observe that the average sign prediction diminishes when the group capitalization increases. This result is evident in Figure (\ref{fig:2}).

When looking at the average total return we detect that for a few stocks, rule returns are negative. However when averaging results for each group, all returns are positive but we cannot affirm that returns are lower when capitalization increases. The same behaviour is appreciated when looking at the Ideal Profit Ratio or Sharpe Ratio.

\begin{table}[htbp]
  \centering
  \caption{Average and standard deviation of selected metrics after 30 independent runs. Stocks within 0-25th. capitalization percentile.}
    \begin{tabular}{rrrrrr}
    \toprule
    \textbf{Ticker} & \textbf{        USIM5} & \textbf{        CPLE6} & \textbf{        BRML3} & \textbf{        RENT3} & \textbf{        ENBR3} \\
    \midrule
    \textbf{Buy \& Hold (train)} & 0.01830 & 0.00499 & -0.01363 & 0.05550 & -0.00081 \\
    \textbf{Return rule (train)} & 0.04794 & 0.01800 & 0.02714 & 0.02624 & 0.01377 \\
    \textbf{Std. Dev} & 0.01274 & 0.02612 & 0.01884 & 0.03378 & 0.01027 \\
    \textbf{Buy \& Hold (test)} & -0.00460 & -0.01504 & -0.00081 & 0.00299 & -0.01888 \\
    \textbf{Sign Prediction} & 73.58\% & 57.13\% & 59.75\% & 54.34\% & 65.29\% \\
    \textbf{Std. Dev} & 0.00098 & 0.01171 & 0.05554 & 0.04627 & 0.11498 \\
    \textbf{Total Return} & 0.03354 & 0.01453 & 0.01142 & -0.03183 & 0.01088 \\
    \textbf{Std. Dev} & 0.01096 & 0.02375 & 0.03158 & 0.03352 & 0.02149 \\
    \textbf{Ideal Profit Ratio} & 0.04491 & 0.02990 & 0.02116 & -0.05172 & 0.02575 \\
    \textbf{Std. Dev} & 0.01096 & 0.04888 & 0.05855 & 0.05445 & 0.05087 \\
    \textbf{Sharpe Ratio} & 3.05859 & 0.61174 & 0.36150 & -0.94981 & 0.50614 \\
    \textbf{Range} & +/-0.0165 & +/-0.005 & +/-0.0083 & +/-0.00516 & +/-0.0116 \\
    \textbf{Std. Dev} & 0.00709 & 0.00000 & 0.00479 & 0.00091 & 0.00531 \\
    \bottomrule
    \end{tabular}%
  \label{tab:q4}%
\end{table}%

\begin{table}[htbp]
  \centering
  \caption{Average and standard deviation of selected metrics after 30 independent runs. Stocks within 25th.-50th capitalization percentile. }
    \begin{tabular}{rrrrrr}
    \toprule
    \textbf{Ticker} &         FIBR3 &         PCAR4 &         NATU3 &         GGBR4 &         CSAN3 \\
    \midrule
    \textbf{Buy \& Hold (train)} & 0.00485 & -0.03741 & 0.01911 & 0.00155 & 0.02732 \\
    \textbf{Return rule (train)} & 0.01939 & 0.04437 & 0.04626 & 0.04216 & 0.00613 \\
    \textbf{Std. Dev} & 0.02666 & 0.00719 & 0.07131 & 0.02489 & 0.01250 \\
    \textbf{Buy \& Hold (test)} & -0.00728 & -0.04381 & 0.00095 & -0.01875 & -0.02629 \\
    \textbf{Sign Prediction} & 54.88\% & 58.81\% & 63.62\% & 67.83\% & 60.73\% \\
    \textbf{Std. Dev} & 0.02602 & 0.02307 & 0.05573 & 0.05273 & 0.06667 \\
    \textbf{Total Return} & 0.02244 & 0.03985 & 0.02661 & 0.03626 & 0.04307 \\
    \textbf{Std. Dev} & 0.01912 & 0.00828 & 0.04353 & 0.02046 & 0.02895 \\
    \textbf{Ideal Profit Ratio} & 0.05835 & 0.11278 & 0.08744 & 0.05288 & 0.07499 \\
    \textbf{Std. Dev} & 0.04972 & 0.02342 & 0.14303 & 0.02985 & 0.05041 \\
    \textbf{Sharpe Ratio} & 1.17350 & 4.81531 & 0.61135 & 1.77197 & 1.48775 \\
    \textbf{Range} & +/-0.01016 & +/-0.0073 & +/-0.005 & +/-0.008 & +/-0.00516 \\
    \textbf{Std. Dev} & 0.00713 & 0.00341 & 0.00000 & 0.00428 & 0.00091 \\
    \bottomrule
    \end{tabular}%
  \label{tab:q3}%
\end{table}%

\begin{table}[htbp]
  \centering
  \caption{Average and standard deviation of selected metrics after 30 independent runs. Stocks within 50-75th. capitalization percentile. }
    \begin{tabular}{rrrrrr}
    \toprule
    \textbf{Ticker} &         LAME4 &         TBLE3 &         WEGE3 &         CCRO3 &         BVMF3 \\
    \midrule
    \textbf{Buy \& Hold (train)} & 0.01427 & 0.01129 & 0.00538 & -0.02447 & 0.00174 \\
    \textbf{Return rule (train)} & 0.03965 & 0.00924 & 0.02033 & 0.05391 & 0.03464 \\
    \textbf{Std. Dev} & 0.03617 & 0.00990 & 0.02430 & 0.03173 & 0.03526 \\
    \textbf{Buy \& Hold (test)} & -0.00177 & -0.01947 & -0.01804 & -0.01482 & -0.01575 \\
    \textbf{Sign Prediction} & 59.28\% & 52.67\% & 64.69\% & 65.65\% & 58.72\% \\
    \textbf{Std. Dev} & 0.05310 & 0.04114 & 0.09738 & 0.07825 & 0.06749 \\
    \textbf{Total Return} & 0.00640 & 0.02244 & 0.00766 & 0.01709 & 0.03197 \\
    \textbf{Std. Dev} & 0.01785 & 0.03648 & 0.01459 & 0.00815 & 0.02498 \\
    \textbf{Ideal Profit Ratio} & 0.01475 & 0.03817 & 0.02548 & 0.04915 & 0.06104 \\
    \textbf{Std. Dev} & 0.04113 & 0.06206 & 0.04850 & 0.02344 & 0.04770 \\
    \textbf{Sharpe Ratio} & 0.35855 & 0.61495 & 0.52532 & 2.09646 & 1.27952 \\
    \textbf{Range} & +/-0.0056 & +/-0.00916 & +/-0.00516 & +/-0.00916 & +/-0.0055 \\
    \textbf{Std. Dev} & 0.00217 & 0.00789 & 0.00091 & 0.00475 & 0.00201 \\
    \bottomrule
    \end{tabular}%
  \label{tab:q2}%
\end{table}%

\begin{table}[htbp]
  \centering
  \caption{Average and standard deviation of selected metrics after 30 independent runs. Stocks within 75th.-100th capitalization percentile.}
    \begin{tabular}{rrrrrr}
    \toprule
    \textbf{Ticker} &         VALE3 &        SANB11 &         PETR3 &         ITUB4 & AMVEB \\
    \midrule
    \textbf{Buy \& Hold (train)} & 0.03415 & 0.00757 & 0.04042 & 0.00419 & -0.06319 \\
    \textbf{Return rule (train)} & 0.00501 & 0.06283 & 0.00242 & 0.00252 & 0.65757 \\
    \textbf{Std. Dev} & 0.01122 & 0.04443 & 0.00671 & 0.00365 & 0.51873 \\
    \textbf{Buy \& Hold (test)} & 0.04247 & -0.05064 & -0.03265 & 0.00383 & 0.06991 \\
    \textbf{Sign Prediction} & 54.59\% & 56.42\% & 52.89\% & 54.77\% & 63.63\% \\
    \textbf{Std. Dev} & 0.03665 & 0.06154 & 0.08615 & 0.04399 & 0.05156 \\
    \textbf{Total Return} & -0.02745 & 0.04937 & 0.01683 & -0.00837 & 0.51680 \\
    \textbf{Std. Dev} & 0.02620 & 0.03814 & 0.02491 & 0.01501 & 0.63330 \\
    \textbf{Ideal Profit Ratio} & -0.04481 & 0.07002 & 0.02347 & -0.01688 & 0.02664 \\
    \textbf{Std. Dev} & 0.04278 & 0.05410 & 0.03472 & 0.03028 & 0.03265 \\
    \textbf{Sharpe Ratio} & -1.04739 & 1.29426 & 0.67585 & -0.55764 & 0.81605 \\
    \textbf{Range} & +/-0.005 & +/-0.0065 & +/-0.005 & +/-0.00516 & +/-0.00616 \\
    \textbf{Std. Dev} & 0.00000 & 0.00418 & 0.00000 & 0.00091 & 0.00215 \\
    \bottomrule
    \end{tabular}%
  \label{tab:q1}%
\end{table}%

\section{Conclusions \label{sec:conclusions}}
Stock markets have an important role not only as asset allocation,  but also as a means through which systemic crisis spreads throughout the economy. So analysing whether markets are informationally efficient is relevant. If a rule based on an ANN (as in this paper), can outperform the market, means that information is not immediately conveyed into prices, as the EMH postulates. 
This situation arises some concerns regarding market regulations and policies recommendations.The aim of our research has been to carry out an intraday forecast on stock returns in the Brazilian stock market. Our study of the time series data has detected an important ability of the ANN to predict sign changes. In fact, the ANN outperforms in several cases a simple buy-and-hold strategy. Still,the performance is far from a perfect prediction. The ANN obtains only between -5\% and 11.2\% of the return that would be obtained if the investor had perfect forecast. The forecast ability is greater on stocks with lower capitalizations. However, this characteristic may be of particular importance for investor and policy makers.

According to our results, we can hightlight two conclusions: (i) the EMH is, albeit imperfect, a useful framework to analyse market efficiency (ii) there are some arbitrage opportunities that can be exploited by using some advanced tools as the ANN. Portfolios managed by big financial institutions are of huge magnitude. Consequently, small abnormal returns eventually represent  millions of dollars. As a caveat, we should say that the true profitability of the technical rule suggested by the ANN can be substantially reduced if we consider transaction costs. In future research, it would be interesting to test other variables beside past prices and trends. For example, traded volume and  volatility can be also taken into account in order to increase the performance of the forecast. Additionally, it would be valuable to analyse if the same size effect we appreciate in stocks is also present in other markets.

\bibliographystyle{spphys}

\bibliography{ref-ann}

\end{document}